\documentclass[aps,prl,showpacs,twocolumn]{revtex4}
\usepackage{amssymb}
\usepackage{graphicx}

\begin{document}

\title{The hard-disk fluid revisited}
\author{Hanqing Zhao$^{1,2}$}
\author{Hong Zhao$^{1,3}$}
\email{zhaoh@xmu.edu.cn}
\affiliation{$^{1}$Department of Physics and Institute of Theoretical Physics and
Astrophysics, Xiamen University, Xiamen 361005, Fujian, China\\
$^{2}$Department of Modern Physics, University of Science and Technology of
China, Hefei 230026, Anhui, China\\
$^{3}$Collaborative Innovation Center of Chemistry for Energy Materials,
Xiamen University, Xiamen 361005, Fujian, China}

\begin{abstract}
The hard-disk model plays a role of touchstone for testing and developing
the transport theory. By large scale molecular dynamics simulations of this
model, three important autocorrelation functions, and as a result the
corresponding transport coefficients, i.e., the diffusion constant, the
thermal conductivity and the shear viscosity, are found to deviate
significantly from the predictions of the conventional transport theory
beyond the dilute limit. To improve the theory, we consider both the kinetic
process and the hydrodynamic process in the whole time range, rather than
each process in a seperated time scale as the conventional transport theory
does. With this consideration, a unified and coherent expression free of any
fitting parameters is derived succesfully in the case of the velocity
autocorrelation function, and its superiority to the conventional
`piecewise' formula is shown. This expression applies to the whole time
range and up to moderate densities, and thus bridges the kinetics and
hydrodynamics approaches in a self-consistent manner.
\end{abstract}

\date{\today }
\maketitle

For a system with translation invariance, the transport theory predicts that
the autocorrelation function (ACF) of a physical quantity, denoted by $C(t)$%
, generally decays as~\cite{theorySP, Lebowitz, Leener, Alder, Alder2,
Alder3, wain-tlnt, Dorfman1, Dorfman2, Dorfman3, Ernst, tlnt, cv-longtail,
review-l-t-coupling, high-density} 
\begin{equation}
\frac{C(t)}{C(0)}=\left\{ 
\begin{array}{ll}
e^{-\xi t}, & \text{kinetics stage;} \\ 
b_{h}t^{-d/2}, & \text{hydrodynamics stage.}%
\end{array}%
\right.
\end{equation}%
Here $t$ is the correlation time, $\xi $ is a characterizing constant, $d$
is the dimension of the system, $b_{h}$ is the amplitude of the power-law
decay function. Given $C(t)$, the transport coefficient of the corresponding
physical quantity can be obtained with the Green-Kubo formula~\cite%
{theorySP, Andrieux, Kubo}. The most important ACFs are those of the
velocity, the energy current and the viscosity current, which will be
referred to as the VACF, the EACF and the VisACF in the following.

Nevertheless, the transport theory has not been fully established. On one
hand, the theoretical predictions have not been well verified yet. Though a
lot of numerical studies have been done since the 1970's~\cite{Alder,
Alder2, wain-tlnt, formula-2d, 2d-tailnotimport, Erpenbeck}, the results are
not accurate enough to conclude until 2008~\cite{pre2008}, Isobe computed
the VACF of the two-dimensional hard-disk fluid and found that the tail is
of the power-law $\sim t^{-1}$ at low densities but logarithmic $\sim (t%
\sqrt{\ln t})^{-1}$ at moderate densities. It implies that the power-law
decay prediction may not be always correct. The logarithmic decay agrees
with the self-consistent mode coupling prediction~\cite{wain-tlnt, tlnt}.
However, the transition form $\sim t^{-1}$ to $\sim (t\sqrt{\ln t})^{-1}$
has not been characterized. For the EACF and VisACF, up to now simulation
results are rare and those allow for drawing a conclusion still lack. On the
other hand, dividing the time dependence of $C(t)$ into the kinetics and
hydrodynamics stages is an expedient measure. To quantitatively calculate
the transport coefficient, a coherent and unified expression is
indispensable. Particularly, for a two dimensional system the power-law tail
of $C(t)$ makes the transport coefficient diverge in the thermodynamical
limit, but for a real system, the measured coefficient should be finite. To
predict the coefficient theoretically, one needs to evaluate the influence
of the long-time tail to reveal when it can be ignored comparing with the
kinetic contribution~and when it becomes dominant\cite{formula-2d,
2d-tailnotimport}. This requires to know the crossover time from the
kinetics stage to the hydrodynamics stage. Numerically, parameter fitting~%
\cite{formula-2d, pre2015} may allow one to construct a unified $C(t)$ 
\textit{within} the time period investigated, but it is risky to extend it
out or to use it in other parameter regimes.

In this work we revisit the hard-disk fluid. First, by large scale
simulations, we calculate the three ACFs and show their deviations from the
theoretical predictions. We then derive a unified expression for the VACF.
The model consists of $N$ disks of unitary mass $m=1$ moving in an $%
L_{x}\times L_{y}$ rectangular area with the periodic boundary conditions.
The system is evolved with the event-driven algorithm~\cite{Alder, cal} at
the dimensionless temperature $T=1$ (the Boltzmann constant is set to be $%
k_{B}=1$). The disk number density is fixing at $n=N/(L_{x}L_{y})=0.01$
throughout and the disk diameter, $\sigma $, is adopted to control the
packing density $\phi =n\pi \sigma ^{2}/4$ (referred to as the density for
short in the following). Three cases, $\sigma =2$, $4$, and $6$
corresponding to $\phi \approx 0.03$, $0.13$, and $0.28$, respectively, are
studied intensively. As a reference, the crystallization density is $\phi
=0.71$, hence our study covers the moderate density regime. Applying the
Enskog formula with the first Sonine polynomial approximation~\cite%
{formula-2d, Gass, pl-eskog}, the diffusion coefficient $D$, the thermal
conductivity $\lambda $, the sheer viscosity $\eta $, and the sound speed $%
u_{s}$ are, respectively, $D=13.4$, $5.70$, $2.76$, $\lambda =0.59$, $0.35$
, $0.36$, $\eta =0.14$, $0.077$, $0.063$, and $u_{s}=1.5$, $1.8$, $2.7$, for
the three densities.

The VACF is defined as $C_{u}(t)=$ $\langle u_{x}(t)u_{x}(0)\rangle $, where 
$u_{x}(t)$ is the $x$-component of the velocity of a tagged disk. Figure
1(a)-(c) show the simulated results obtained with $10^{10}$ ensemble samples
for $L_{x}=L_{y}=2000$ ($N=40000$). The time range free from the finite-size
effects is $0\leq t<t_{f}=L_{x}/(2u_{s})$~\cite{pre2008, chenfinite,SM}. For
the three densities, $t_{f}=667$, $556$, and $370$, respectively. In this
time range, the initial exponential decaying stage and the long-time tail
can be observed in all the three cases. The hydrodynamics prediction $%
C_{u}(t)/C_{u}(0)=[8\pi (D+\nu )n]^{-1}t^{-1}$ \cite{Alder, Alder2, Alder3,
wain-tlnt, Dorfman1, Dorfman2, Dorfman3} is also plotted for comparison,
where $\nu =\eta m/n$ is the viscosity diffusivity. It can be seen that the
predicted $\sim t^{-1}$ tail is close to the simulation result, but as the
density increases, the deviation grows. The diffusion coefficients
calculated following the Green-Kubo formula, $D(t)=\int_{0}^{t}C_{u}(t^{%
\prime })dt^{\prime }$, are shown in Fig.~1(d).

\begin{figure}[tbp]
\centering
\includegraphics[width=8.5cm]{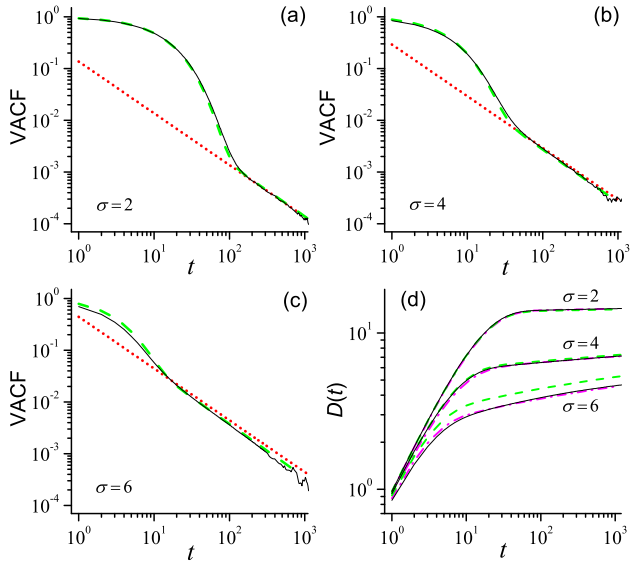}
\caption{(Color online) (a)-(c) The VACFs obtained by, respectively,
simulations (black solid lines), the unified formula Eq.~(5) (green dashed
lines), and the hydrodynamics theory (red dotted lines). (d) The diffusion
coefficients calculated with the VACFs obtained by simulations (black solid
lines) and by the unified formula Eq.~(5). For the latter, the green dashed
lines are for the results with the kinetic transport coefficients given by
the Enskog equation and the magenta dash-dotted lines are for the results
with the corrected coefficients.}
\label{fig1}
\end{figure}

Calculating the EACF $C_{J}(t)=\langle J_{x}(t)J_{x}(0)\rangle $ is $N$
times harder than calculating the VACF. Here $J_{x}(t)=\sum_{i}j_{x}^{i}(t)$%
, where $j_{x}^{i}(t)=|\mathbf{v}^{i}|^{2}v_{x}^{i}$ is the $x$-component of
the energy current of the $i$th disk. For a given simulation run we can
obtain $N$ ensemble samples for calculating $C_{u}(t)$ as every disk can be
taken as the tagged disk but only one for calculating $C_{J}(t)$ because the
total current, $J_{x}$, involves the contributions of all the disks~\cite%
{wain-tlnt,tlnt}. This is the reason why the hydrodynamics prediction of the
EACF has not been conclusively tested. To decrease the simulation difficulty
we consider a smaller size, i.e., $L_{x}\times L_{y}=3000\times 400$ $%
(N=12000)$. Correspondingly, $t_{f}=1000$, $833$, and $535$ for the three
densities. Figure 2(a)-(c) show the results of $C_{J}(t)$ calculated with $%
10^{9}$ ensemble samples. For $\sigma =2$, a perfect $\sim t^{-1}$ tail is
observed, but the value of $C_{J}(t)$ at the tail is one time larger than
the hydrodynamics prediction~\cite{Alder, Alder2, Alder3, wain-tlnt,
Dorfman1,Dorfman2, Dorfman3} that $C_{J}(t)/C_{J}(0)=[4\pi (\eta /m+\lambda
/2k_{B})]^{-1}t^{-1}$. For $\sigma =4$ and $6$, the EACF shows a multistage
decaying behavior -- after the initial exponential decaying stage there
appears another fast decaying stage, before a power-law tail slower than $%
\sim t^{-1}$ follows. Figure 2(d) shows the corresponding thermal
conductivity calculated following the Green-Kubo formula $\lambda (t)=\frac{1%
}{k_{B}T^{2}L_{x}L_{y}}\int_{0}^{t}C_{J}(t^{\prime })dt^{\prime }$.

\begin{figure}[tbp]
\centering
\includegraphics[width=8.5cm]{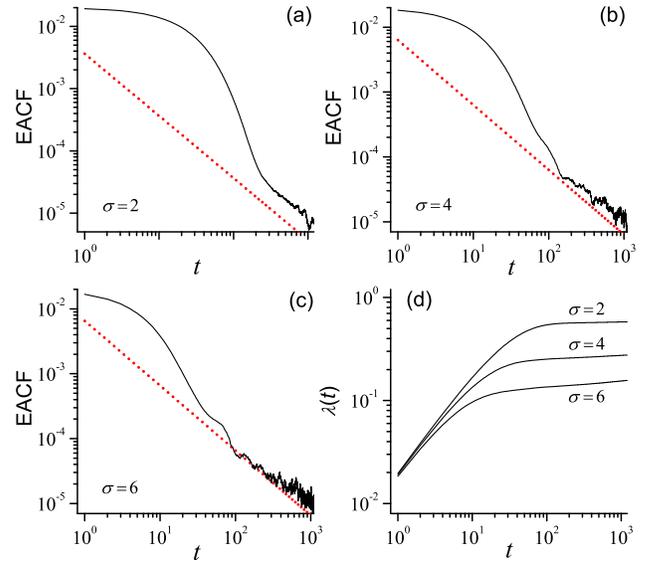}
\caption{(Color online) (a)-(c) The EACFs obtained by simulations (black
solid lines) and by the hydrodynamics theory (red dotted lines). (d) The
thermal conductivity calculated based on the EACFs obtained by simulations.}
\label{fig2}
\end{figure}

Calculating the VisACF $C_{vis}(t)=\langle J_{vis}(t)J_{vis}(0)\rangle $
suffers from the same difficulty. Here $J_{vis}(t)=%
\sum_{i}u_{x}^{i}(t)u_{y}^{i}(t)$. Taking $L_{x}=L_{y}=1000$ $(N=10^{4})$
and $\sigma =4$, we show in Fig.~3(a) the VisACF calculated with $10^{10}$
ensemble samples. Though for $t<t_{f}=278$ the VisACF decays fast for
several orders, it is still uncertain if a power-law tail follows. Indeed,
the VisACF may drop to be negative from $t\approx 70$ to $100$. The demanded
huge amount of samples make the computation so difficult that we can only
provide the results for one case ($\sigma =4$) as an example. The
hydrodynamics prediction $C_{vis}(t)/C_{vis}(0)=(32\pi )^{-1}[m/\eta +(\eta
/m+\lambda /2k_{B})^{-1}]t^{-1}$~\cite{Alder, Alder2, Alder3, wain-tlnt,
Dorfman1, Dorfman2, Dorfman3} is also plotted for comparison. Figure~3(b)
shows the shear viscosity by following the Green-Kubo formula $\eta
(t)=m\int_{0}^{t}C_{vis}(t^{\prime })dt^{\prime }$.

\begin{figure}[tbp]
\centering
\center\includegraphics[width=8.5cm]{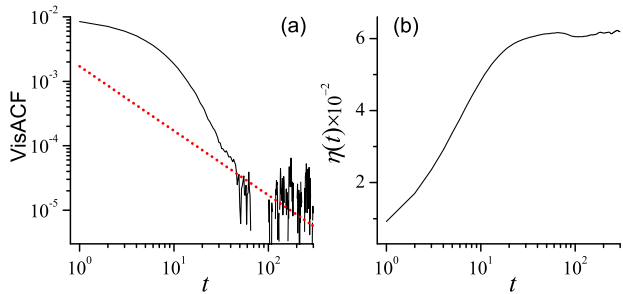}
\caption{(Color online) (a) The simulation result of the VisACF and (b) the
corresponding sheer viscosity for $\protect\sigma =4$.}
\label{fig3}
\end{figure}

Next, we derive the unified expression for the VACF. Our key consideration
is that the VACF is governed by two physical processes simultaneously. One
is collisions of the tagged disk with other surrounding disks, referred to
as the kinetic process, through which $C_{u}(0)$ will be transferred to
other disks from the tagged disk. Let $C_{u}^{k}(t)$ denote the portion of $%
C_{u}(0)$ that has not been transferred at time $t$. Following the kinetics
theory, it decays exponentially: $C_{u}^{k}(t)=C_{u}(0)\exp ({-\frac{k_{B}T}{%
mD^{k}}t})$~\cite{Alder, Alder2, Alder3, Dorfman1, Dorfman2, Dorfman3},
where $D^{k}$ is the kinetic diffusion constant. Meanwhile, it is possible
for the transferred portion to feedback to the tagged disk by ring collisions%
\cite{Alder2}, which is referred to as the hydrodynamic process. The amount
returns to the tagged disk at time $t$, denoted by $C_{u}^{h}(t)$,
contributes to the hydrodynamics diffusion constant $D^{h}$. The VACF is
thus a sum of these two portions, 
\begin{equation}
C_{u}(t)=C_{u}^{k}(t)+C_{u}^{h}(t),
\end{equation}%
and the diffusion constant is divided into the kinetic and hydrodynamic
parts as $D=D^{k}+D^{h}$ accordingly. To obtain $C_{u}^{h}(t)$, it is
necessary to investigate how $C_{u}(0)$ is delivered to the surroundings.
This reduces to investigating the relaxation process of the momentum $%
\mathbf{p}_{c}=m\mathbf{u}_{c}$ initially carried by the tagged disk, which
can be approached with the spatiotemporal correlation function~\cite%
{zhao2006,chendiffusion} 
\begin{equation}
c(\mathbf{r},t)=\frac{\langle \mathbf{p}_{c}\cdot \mathbf{p}(\mathbf{r}%
,t)\rangle }{\langle |\mathbf{p}_{c}|^{2}\rangle }+\frac{n}{N-1}.
\end{equation}%
Here $\mathbf{p}(\mathbf{r},t)$ is the momentum density of the system. It is
found that $c(\mathbf{r},t)$ is axisymmetric with respect to $\mathbf{r}=0$;
it has one center peak surrounded by a `crater' (see Fig.~4(a)-(b) for the
intersection of $c(\mathbf{r},t)$ with $y=0$) and the center peak can be
well fitted by the Gaussian function $c^{center}(\mathbf{r},t)=\frac{a_{\nu }%
}{4\pi \widetilde{\nu }t}\exp ({-\frac{r^{2}}{4\widetilde{\nu }t}})$ with $%
a_{\nu }=1/2$ and $\widetilde{\nu }=14.3$, $8.3$, and $7.9$ for $\sigma =2$, 
$4$, and $6$, respectively. The function $c(\mathbf{r},t)$ gives the portion
of $C_{u}(0)$, that transfers to a unit area centering $\mathbf{r}$ at time $%
t$. There are $n$ disks on average in this area, and each of them carries a
portion, i.e., $c(\mathbf{r},t)/n$, of $C_{u}(0)$. Suppose that the tagged
disk appears in this area with the probability $\rho (\mathbf{r},t)$, then
on average the portion of $C_{u}(0)$ it carries is $\rho (\mathbf{r},t)c(%
\mathbf{r},t)/n$. The total amount carried by it is therefore $%
C_{u}^{h}(t)=(1/n)C_{u}(0)\int c(\mathbf{r},t)\rho (\mathbf{r},t)d\mathbf{r}$%
.

\begin{figure}[tbp]
\centering
\includegraphics[width=8.5cm]{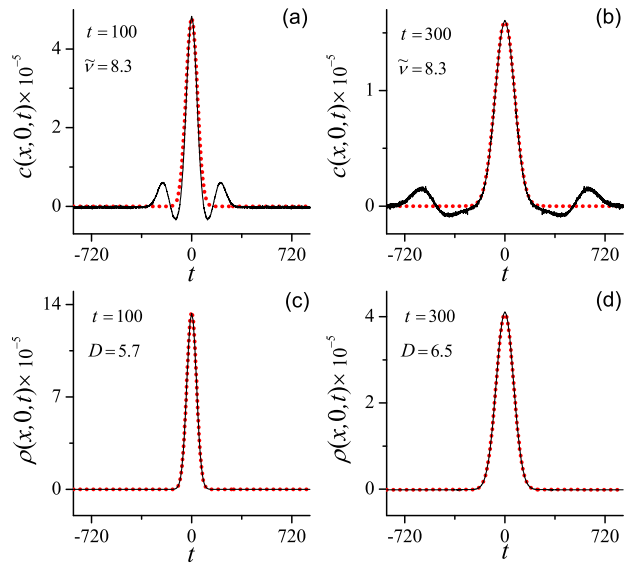}
\caption{(Color online) (a)-(b) The intersection of $c(\mathbf{r},t)$ and
(c)-(d) the intersection of $\protect\rho (\mathbf{r},t)$ at $t=100$ and $%
300 $ (black solid lines). The red dotted line in each panel is the best
Gaussian fitting to the center peak of $c(\mathbf{r},t)$ or $\protect\rho (%
\mathbf{r},t)$. $\protect\sigma =4$.}
\label{fig4}
\end{figure}

The probability function $\rho (\mathbf{r},t)$ can be measured directly by
tracing the tagged disk. It is found to overlap perfectly with a Gaussian
function (see Fig.~4(c)-(d) for its intersection) with a time dependent
diffusion coefficient, i.e., $\rho (\mathbf{r},t)=\frac{1}{4\pi D(t)t}\exp ({%
-\frac{r^{2}}{4D(t)t}})$. As $\rho (\mathbf{r},t)$ decays exponentially as $%
r^{2}$, we can replace $c(\mathbf{r},t)$ with $c^{center}(\mathbf{r},t)$ in
the integrand for calculating $C_{u}^{h}(t)$. With this simplification we
have 
\begin{equation}
C_{u}^{h}(t)/C_{u}(0)=a_{\nu }[4\pi (D(t)+\widetilde{\nu })n]^{-1}t^{-1}.
\end{equation}

Noting that Fig.~4 represents the case that the hydrodynamics effects become
completely dominant. For $t>100$, $C_{u}^{k}(t)$ has decayed to a negligibly
small value($<10^{-8}$ with $D^{k}=5.70$), implying that $C_{u}(0)$ has
transferred to the surrounding disks almost completely. Equation~(4) thus
characterizes the situation at large times. At short times, the portion of $%
C_{u}(0)$ the hydrodynamics process accounts for is $C_{u}(0)[1-\exp (-\frac{%
k_{B}T}{mD^{k}}t)]$; Assuming that $c(\mathbf{r},t)$ for this portion has
the same structure as shown in Fig. 4(a)-4(b), it is straightforward to have 
\begin{equation}
C_{u}^{h}(t)/C_{u}(0)=a_{\nu }[1-\exp (-\frac{k_{B}T}{mD^{k}}t)][4\pi (D(t)+%
\widetilde{\nu })n]^{-1}t^{-1}.
\end{equation}%
This extended expression applies in both the kinetics and the hydrodynamics
stages.

The parameter $a_{v}$ and $\widetilde{\nu }$ can be connected to the
properties of the hydrodynamic modes analytically ~\cite{SM}. Let $n(\mathbf{%
r},t)$ and $\mathbf{u}(\mathbf{r},t)$ be the disk number density and the
velocity density, we have $\mathbf{p}(\mathbf{r},t)=mn(\mathbf{r},t)\mathbf{u%
}(\mathbf{r},t)=m\mathbf{j}(\mathbf{r},t)$, considering the hydrodynamics
assumption~\cite{theorySP} that local deviations of hydrodynamic variables
from their average values are small. Here $\mathbf{j}(\mathbf{r},t)$ is the
local disk current. Applying the hydrodynamics analysis to solve the
linearized conservation laws for the disk number, the energy, and the
momentum with initial conditions of $\delta $-function impulses of $\delta (%
\mathbf{r})\Delta n$, $\delta (\mathbf{r})\Delta T$ and $\delta (\mathbf{r})%
\mathbf{p}_{c}$, we obtain $c(\mathbf{k},t)=(k_{y}^{2}/k^{2})\exp (-\nu
k^{2}t)$ in the wave-vector space, where $\Delta n$ and $\Delta T$ represent
the deviation of the disk density and the temperature induced by the tagged
particle, and $\mathbf{k}$ is the wave vector in the Fourier space. With a
rough estimation of $k_{y}^{2}/k^{2}\sim 1/2$, it appears $c(\mathbf{k}%
,t)=(1/2)\exp (-\nu k^{2}t)$ and gives $c(\mathbf{r},t)=(1/8\pi \nu t)\exp
(-r^{2}/4\nu t)$ in the real space. The expression of $c(\mathbf{r},t)$
implies $a_{v}=1/2$ and $\widetilde{\nu }=\nu $. With this connection,
Eq.~(4) is exactly the same as that of the hydrodynamics theory.

In principle, $\nu $ is time-dependent according to the hydrodynamics
theory, but based on our numerical observation of the relaxation of $%
c^{center}(\mathbf{r},t)$ and the fact that $\eta $ converges in time [see
Fig.~3(d)], it can be assumed to be a time-independent constant up to
moderate densities. Previous numerical studies using the Helfand-Einstein
formula have also shown that the shear viscosity does not depend on the
system size either~\cite{formula-2d, 2d-tailnotimport}, which supporting the
constant $\nu $ assumption as well.

Inserting $C_{u}^{h}(t)$ into the Green-Kubo formula, we have 
\begin{equation}
D^{h}(t)=\int_{0}^{t}C_{u}^{h}(t^{\prime })dt^{\prime }.
\end{equation}%
It is interesting to note that the self-consistent solutions~\cite%
{wain-tlnt, tlnt}, i.e., $C_{u}^{h}(t)/C_{u}(0)=\sqrt{1/16\pi n}(t\sqrt{\ln
(t)})^{-1}$ and $D^{h}(t)=\sqrt{k_{B}T\ln (t)/4\pi mn}$, are asymptotic
solutions of Eq.~(5) and (6) in the long-time limit where $D^{h}(t)\gg
D^{k}+\nu $, i.e., $t>\exp [(4\pi mn/k_{B}T)(D^{k}+\nu )^{2}]$. Using the
Enskog results of $D^{k}$ and $\nu $, it can be estimated that this time
scale is about $10^{23}$, $10^{10}$, and $10^{4}$, respectively, for $\sigma
=2$, $4$, and $6$. During the transition process which may contribute a
dominant part to the diffusion constant, the self-consistent asymptotic
solutions are not exact.

In order to solve the coupled equations (5) and (6) accurately, we turn to
the iterative algorithm: We set $D^{h}(t)=0$ as the first trial solution and
substitute it into Eq.~(5) to get $C_{u}^{h}(t)$, then put it into Eq.~(6)
to get the next trial solution of $D^{h}(t)$, and so on. In general if $%
D^{h}(t)$ increases, $C_{u}^{h}(t)$ will decrease and make $D^{h}(t)$
decrease, and vice versus, hence the convergence of the iteration is
guaranteed. Indeed, usually the solutions converge after only several
iterations. The predicted VACFs [Fig.1(a)-(c)] and the corresponding
diffusion coefficients [Fig.1(d)] agree with the simulation results quite
well, except that the diffusion coefficient $D(t)$ show a shift from the
simulation result as the density increases.

This deviation should be induced by the inaccuracy of the kinetics transport
coefficients that we have employed. With the simulation data of $C_{u}(t)$,
we can estimate the kinetics diffusion constant $D^{k}$. The kinetics
process plays a role mainly before the time, denoted as $\tau $, at which $%
C_{u}(t)$ turns from the exponential decay to the followed tail. For
example, for $\sigma =2$, $\tau \approx 110$, at which $C_{u}^{k}$ has
decayed to $C_{u}^{k}(\tau )/C_{u}(0)\sim 10^{-4}$. Truncating the
Green-Kubo integration at $\tau $, we have $D(\tau )=13.97$, $6.40$, and $%
2.85$ for $\sigma =2$, $4$, and $6$, respectively. These values can be
considered as the upper bound of $D^{k}$. Subtracting $D^{h}(\tau )$ from
it, we get the estimated $D^{k}$; i.e., $D^{k}=13.61$, $5.50$, and $2.35$,
correspondingly. These values are close to the Enskog approximations, within
maximally $13\%$ errors. Meanwhile, the relation $\widetilde{\nu }=\nu $ is
also not accurate; it is a result of ignoring the anisotropic feature of the
momentum diffusion \cite{SM}. More properly, $\widetilde{\nu }$ measured by
the direct simulation should be employed to characterize the momentum
diffusion instead of $\nu $. With these corrections of $D^{k}$ and $\nu $,
the shift of $D(t)$ can be well suppressed [see Fig.~1(d)].

Similarly, we can estimate the upper bounds for the heat conductivity. From
Fig.~2(d) we have the heat conductivity $\lambda \leq 0.580$, $0.258$, and $%
0.139$ for $\sigma =2$, $4$, and $6$, respectively, which deviates at least $%
2\%$, $36\%$, and $160\%$ from the Enskog approximations. Therefore, the
Enskog equation is relatively precise for the kinetic diffusion constant,
but lacks accuracy for the heat conductivity and the sheer viscosity at
higher densities.

With the unified expression of $C_{u}(t)$, we can estimate the hydrodynamics
contribution to the diffusion constant to systems of macroscopic sizes. For
example, the average distance between two neighboring molecules in the air
is about $10^{-9}$ meter, implying that if our model has a macroscopic size,
say one centimeter, we have $L_{x}$, $L_{y}\sim 10^{8}$. For such a size,
the time a disk diffuses freely without being influenced by the boundaries
is $t\sim L_{x}/(2u_{s})\sim 10^{7}$. Taking this time as the truncation
time of integration in Eqs.~(5) and (6), our iteration algorithm gives $%
D^{h}(t)/D^{k}\approx 0.15$, $0.5$, and $10$ for $\sigma =2$, $4$,and $6$,
respectively, suggesting that in a dilute system it is the kinetics
contribution that dominates, but as the density increases, the hydrodynamics
contribution increases dramatically and the kinetics contribution turns to
be negligible.

In summary, beyond the dilute limit, the accuracy of the hydrodynamics
theory is not sufficient in describing the ACFs at least in the transient
stage that is essential for calculating the transport coefficients. For the
VACF, the numerically observed tail is between $\sim t^{-1}$ and $\sim (t%
\sqrt{\ln (t)})^{-1}$. For the EACF, we have evidenced the power-law tail
but the exponent agrees with the hydrodynamics prediction only in a very
dilute system. As the density increase, a multistage decaying phenomenon is
observed, and the long-time tail is slower than $\sim t^{-1}$. The VisACF
decays much faster than the VACF and the EACF. The long-time tail has not
been observed in our example. In addition, we have estimated the upper
bounds of transport coefficients using the simulated ACFs, and reveal that
the Enskog equation generally used for approximating the kinetics transport
coefficients need be improved particularly at higher densities.

For the VACF, our intuitive representation of the ring-collision mechanism
and the iterative algorithm lead us to a unified and coherent expression
valid in the whole time range. The key point is to distinguish the kinetics
and the hydrodynamics processes and investigate them respectively over the
whole time range. Particularly, we emphasize that the hydrodynamic
contribution at short times should not be ignored. This is different from
the traditional treatments that divide the relax process into separated
stages. Extension of our method to the EACF and VisACF is open.

\textbf{Acknowledgments}

Very useful discussions with J. Wang, Y. Zhang and Dahai He are gratefully
acknowledged. This work is supported by the National Natural Science
Foundation of China (Grant No. 11335006), and the NSCC-I computer system of
China.

\end{document}